\documentclass[aps,prl,preprint]{revtex4-1}

\usepackage{physics}
\usepackage{graphicx}
\usepackage{siunitx}
\usepackage{verbatim}
\usepackage{bm}

\begin{document}

%Title of paper
\title[]{An experimental investigation of measurement-induced disturbance and time symmetry in quantum physics}
\author{D. Curic$^1$, M.C. Richardson$^1$, G.S. Thekkadath$^2$, J. Fl\'orez$^1$, L. Giner$^1$, J.S. Lundeen$^1$}

\address{$^1$Department of Physics and Centre for Research in Photonics,
University of Ottawa, 25 Templeton Street, Ottawa, Ontario, K1N 6N5, Canada}
\address{$^2$Clarendon Laboratory, University of Oxford, Parks Road, Oxford, OX1 3PU, UK}

\begin{abstract}
    Unlike regular time evolution governed by the Schr\"odinger equation, standard quantum measurement appears to violate time-reversal symmetry. Measurement creates random disturbances (e.g., collapse) that prevents back-tracing the quantum state of the system. The effect of these disturbances is explicit in the results of subsequent measurements. In this way, the joint result of sequences of measurements depends on the order in time in which those measurements are performed. One might expect that if the disturbance could be eliminated this time-ordering dependence would vanish. Following a recent theoretical proposal [A. Bednorz et al 2013 New J. Phys. \pmb{15} 023043], we experimentally investigate this dependence for a kind of measurement that creates an arbitrarily small disturbance, weak measurement. We perform various sequences of a set of polarization weak measurements on photons. We experimentally demonstrate that, although the weak measurements are minimally disturbing, their time-ordering affects the outcome of the measurement sequence for quantum systems.
\end{abstract}

\maketitle

A fundamental open question in physics is the role of time in quantum
mechanics~\cite{muga2007time,hilgevoord2005time,Horsman20170395}. 
While observables such as position and momentum are represented by operators, time in the Schr\"odinger equation appears only as ordinary number-parameter, just as in classical mechanics~\cite{hilgevoord2005time}. In view of relativistic theories of physics which famously treat time and space on equal footing, this distinction is problematic. In fact, while Heisenberg's uncertainty principle between energy and time appears to suggest that a time operator conjugate to the total energy operator exists, attempts to create such an operator lead to contradictions~\cite{pauli1958encylopaedia}. Similar issues confound attempts to create an observable for the time it takes for a particle to tunnel through a potential barrier, or even
the time of arrival of a particle at a detector~\cite{hauge1989tunneling,steinberg1993measurement,muga2000arrival}. Another issue is that it is widely believed that information is conserved in quantum physics. This follows from the unitary time-reversible evolution in the Schr\"odinger equation. Yet, it is possible that the passage and direction of time might be set and discerned only by sequences of `events'~\cite{PhysRevD.92.045033}. The only recordable `events' in quantum
mechanics are the results of measurements~\cite{aharonov1964time}, which themselves are not unitary, time-reversible, or conserve information. In this work, we experimentally investigate sequential measurements on quantum systems in order to gain insight into the role that time plays in the theory.  

Specifically, we test whether the results (i.e., `events') of a sequence of measurements depends on the order in time in which they are performed. In both classical and quantum physics, measurements may induce a disturbance~\cite{bednorz2013noninvasiveness}. Since the disturbance will affect subsequent measurements, the results of sequences of measurements may be dependent on the order in time (i.e., time-ordering) in which they are performed. In quantum physics, this is particularly apparent for sequences of incompatible observables, those that do not commute. However, this disturbance can be arbitrarily reduced, at the expense of information gain per measurement trial. Such minimally disturbing measurements are often referred to as weak measurement. Given that the disturbance was the source of the time-ordering dependence, one might expect the time-ordering dependence to disappear for such minimally disturbing measurements. 

Minimally-disturbing measurements have proven useful for probing quantum systems~\cite{aharonov1988result,thekkadath2016directly,lundeen2011direct,salvail2012full,kim2012protecting,ahnert2004weak}. They have also recently attracted wide interest as a signal amplification technique and for studying paradoxes in quantum physics~\cite{aharonov1988result, lundeen2009experimental}. Sequences of weak measurements and continuous weak measurements have also become important tools for exploring features of quantum mechanics that are impossible to study with conventional methods~\cite{PhysRevA.76.062105,PhysRevLett.93.163602,PhysRevLett.119.220507}. It has been shown that the joint result of a sequence of weak measurements is invariant to time-orderings in classical physics~\cite{bednorz2013noninvasiveness,franke2012time}, as expected. Particularly relevant for this work, in quantum physics, sequences of weak measurements can be used for simultaneously measuring incompatible observables~\cite{thekkadath2016directly, piacentini2016measuring}. Although such weak measurements are minimally disturbing, the results are, surprisingly, predicted to vary with time-ordering, in contrast to the classical case. 

We seek to experimentally demonstrate these time-ordering effects and investigate why they occur. We use the polarization of photons as our quantum system, and thus weakly measure incompatible observables, such as horizontal and diagonal polarization projectors. We show that for such observables the order in which weak measurements are made matters. We study this effect for sequences of two and three measurements. We also probe the role of quantum coherence in time-ordering by testing the case in which the state is incoherently polarized.

The notion of weak measurement naturally emerges from a general model for measurement known as the Von Neumann or indirect measurement model~\cite{wiseman2009quantum}. Almost all
measurements, classical or quantum, fit within this model. In it, a measurement apparatus interacts with the measured system, therein disturbing it~\cite{bednorz2013noninvasiveness}. Since the interaction creates the disturbance, an obvious method to reduce the latter is to weakly couple the measurement apparatus to the measured system. The
measurement apparatus consists of a pointer $\mathcal{P}$ whose momentum $\pmb{p}$ is coupled to measured observable $\pmb{A}$ on the measured system $\mathcal{S}$ via unitary interaction 

\begin{equation}
\pmb{U}=\mathrm{exp}\left(-i \delta \pmb{A}\otimes \pmb{p}\right)=\sum_{a}\ket{a}\bra{a}\otimes \pmb{T}(\delta a).\label{VonNeumannEqn1}
\end{equation}
Here, $\delta$ denotes the interaction strength. On the right-side, we
have rewritten $\pmb{U}$ in the eigenbasis of $\pmb{A}$, where $\ket{a}$ is
an eigenstate of $\pmb{A}$ with eigenvalue $a.$ The translation operator, $\pmb{T}( \delta a)=\mathrm{exp}(-i \delta a\pmb{p})$, shifts the pointer's position, i.e., $\pmb{T}( \delta a)\ket{x}=\ket{x- \delta a}$, where $\ket{x}$ is a position eigenstate. If the initial pointer state $\ket{\psi}$ has a position width $\sigma< \delta a$ then the post-interaction pointer position $x= \delta a$ unambiguously indicates the measurement result $a$. This is the case commonly found in conventional measurements, such as a polarizing beam splitter (PBS), or measuring the spin of a silver atom with a Stern-Gerlach apparatus. With this explicit model, one can consider reducing the
interaction strength so that $\delta a\ll\sigma$. Since the pointer now
extends over multiple indicator marks, $\delta a$ (where $a$ is a particular value in the spectrum of $\pmb{A}$), a single trial's measurement result will be ambiguous.  However, averaged over many trials the mean measurement result is proportional to the average pointer position, $\delta \langle \pmb{A}\rangle=\langle \pmb{x} \rangle,$ regardless of measurement strength. It is this average result, the expectation value of $\pmb{A}$, that we will study.

To extend the above formalism to include a sequence of measurements $\pmb{A}_N \cdots \pmb{A}_2 \pmb{A}_1$, one composes a product of unitaries $\pmb{U}_N \dots \pmb{U}_2 \pmb{U}_1$, where $\pmb{U}_i$ takes the form of Eq.~\ref{VonNeumannEqn1}. Here, $\pmb{A_1}$ is the first observable measured and $\pmb{A}_N$ is the last. Thus, each observable $\pmb{A}_i$ in the sequence of measurements is independently coupled to a distinct pointer $\mathcal{P}_i$ with state $\ket{\psi(x_i)}$. The final result of the measurement is the expectation value $\langle \pmb{x}_1^{(\pmb{A}_1)} \pmb{x}_2^{(\pmb{A}_2)} \cdots \pmb{x}_N ^{(\pmb{A}_N)} \rangle/\delta^N$, where the superscript ($\pmb{A}_i$) is the observable to which the $\mathcal{P}_i$ pointer is coupled to~\cite{lundeen2005practical}. With this, we can consider the effect of reducing the interaction strength and testing different time-orderings of the measurements. For example, how does $\langle \pmb{x}_1^{(\pmb{A}_1)} \pmb{x}_2^{(\pmb{A}_2)} \rangle$ compare to $\langle \pmb{x}_1^{(\pmb{A}_2)} \pmb{x}_2^{(\pmb{A}_1)} \rangle$?

\begin{figure}[h!]
    \centering
    \includegraphics[scale=.5]{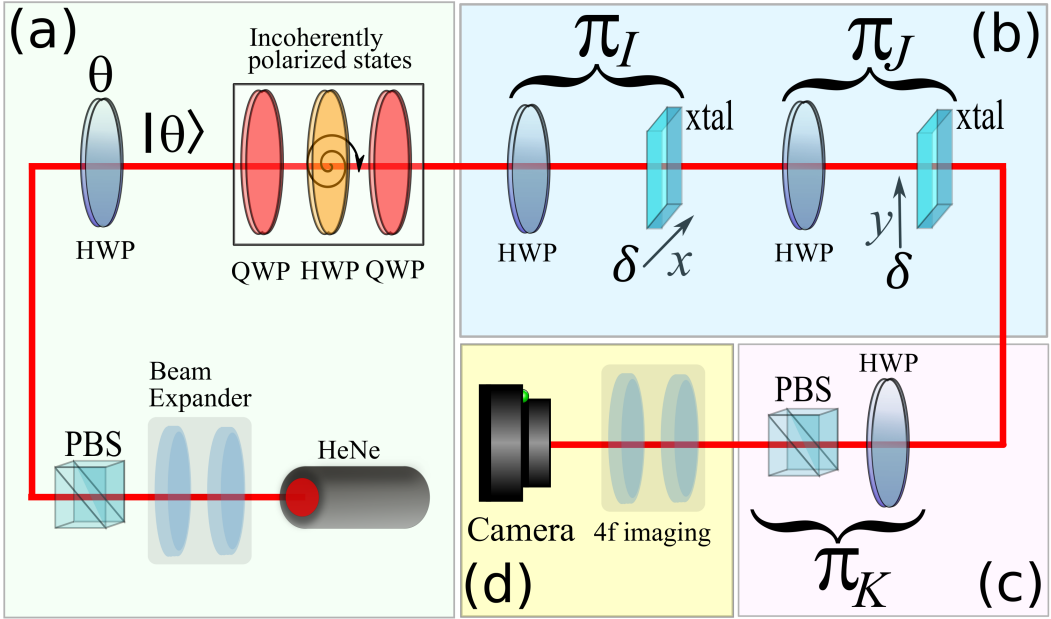}
    \caption{Setup to measure the effect of time-ordering of weak measurements. \pmb{(a)}. State preparation: a beam expander is used to decrease the interaction strength by increasing the width $\sigma$ of the pointer, i.e., the photon transverse distribution. The input system polarization state $\ket{\theta}$ is prepared by a polarizing beam splitter (PBS) and a half-wave plate (HWP). Using quarter-wave plates (QWP) and a HWP (QWP at 45$^{\circ}$ $\rightarrow$ HWP $\rightarrow$ QWP at 90$^{\circ}$) incoherently polarized light can be generated. The HWP is attached to a motor that spins at a rate faster than the collection time of the camera. \pmb{(b)} Weak Measurements: Each weak measurement is implemented by a HWP followed by a walk-off crystal (xtal). The first effects the $\pmb{\pi}_I$ projector by shifting $\ket{I}$ polarized light by $\delta < \sigma$ in the $x$ direction. Likewise, the second implements the $\pmb{\pi}_J$ projector by shifting $\ket{J}$ polarized light by $\delta$ in the $y$ direction. \pmb{(c)} Strong measurement: The combination of a HWP and a PBS realizes the third measurement of $\pmb{\pi}_H$ when desired, and is taken out of the setup otherwise. \pmb{(d)} A 4$f$ system images the shifted beam onto a imaging camera.}
    \label{fig:setup}
\end{figure}

The experimental setup is shown in Fig.~\ref{fig:setup}. It is technically
challenging to use spatially distinct systems as our pointer $\mathcal{P}$ and measured
system $\mathcal{S}$. Instead, we use distinct degrees of freedom of a photon. The
measured system $\mathcal{S}$ is the polarization degree of freedom,
whereas the pointer $\mathcal{P}$ is the photon's transverse position.
A HeNe laser at 633 nm followed by a polarizing beam splitter (PBS) and a rotatable half-wave
plate (HWP), set at an angle $\theta,$ prepares an ensemble of identically polarized photons
as our system input state, $\ket{\theta}=\cos{(2 \theta)}\ket{H}+\sin{(2\theta)
}\ket{V}$.

The coupling between $\mathcal{S}$ and $\mathcal{P}$ can be accomplished
using polarization dependent walk-off in birefringent crystals. The
walk-off transversely shifts the extraordinary polarized photons by
$\delta$ relative to ordinary polarized photons. A sequence of two weak
measurements requires two independent pointers, which we take as the
two transverse spatial degrees of freedom, $x$ and $y$, of the photon,
each with same initial wavefunction: $\braket{x,y}{\psi_x, \psi_y}=\psi(x)\psi(y)=(2\pi\sigma^{2})^{-\frac{1}{2}}\mathrm{exp}\left(-(x^{2}+y^2)/4\sigma^{2}\right)$. A  beam  expander  magnifies
the  HeNe's  transverse  Gaussian  mode to set the  a
width of $\sigma$ = \SI{600}{\micro\meter}. With these two pointers, the measurement coupling is implemented with two walk-off crystals, one of which displaces the horizontal polarization
along the $x$-axis, followed by an identical crystal rotated by $90^{\circ}$
so that it shifts the vertical polarization along the $y$-axis. Both
crystals impart a shift of $\delta =$ \SI{160}{\micro\meter} ensuring
that we are in the weak measurement regime, $\delta/\sigma \approx 0.25.$

In order to change which observable each crystal implements we add
waveplates that effectively rotate the basis of the measurement. With this, we use either the  $\pmb{x}_1 = \pmb{x}$ or $\pmb{x}_2 = \pmb{y}$ positions to read out measurements of the $\pmb{A}_1 = \ket{I}\bra{I} = \pmb{\pi}_I$, and $\pmb{A}_2 = \ket{J}\bra{J} = \pmb{\pi}_J$ polarization projectors, depending on the ordering. We
demonstrate the effect of time-ordering on weak measurements by comparing
the result of a sequential measurement $\langle \pmb{x}^{(\pmb{\pi}_{I})}\pmb{y}^{(\pmb{\pi}_{J})}\rangle$
with the reversed sequence $\langle \pmb{x}^{(\pmb{\pi}_{J})}\pmb{y}^{(\pmb{\pi}_{I})}\rangle$. In all
cases, the expectation values $\langle \pmb{xy}\rangle$ of the photon transverse
two-dimensional distribution are found by imaging onto a camera.

\begin{figure}[h!]
    \centering
    \includegraphics[scale=0.5]{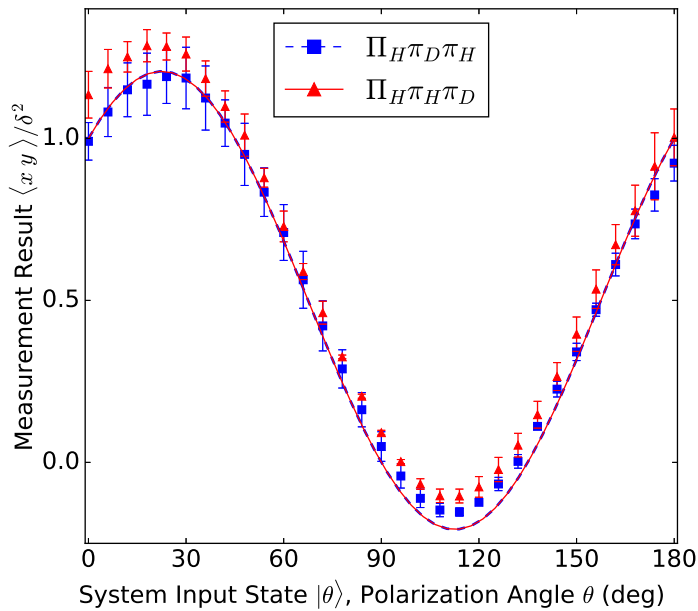}
    \caption{The measurement result $\langle \pmb{xy} \rangle/\delta^2$ for a sequence of two measurements. In one, the sequence $\pmb{\pi}_H \pmb{\pi}_D$ is measured (red triangles), and in the other, $\pmb{\pi}_D \pmb{\pi}_H$ is measured (blue squares). Since the points agree within error, the plot shows that the results do not depend on the order in which the measurements are performed. The red solid and blue dashed line are the respective theoretical curves. Error bars are the standard error obtained by averaging over four experimental runs. Imperfections in the HWP birefringence likely introduces the differences between the experimental points and the theoretical curve, as they can create systematic errors not only when preparing the input polarization state $\ket{\theta},$ but also when aligning the walk-off crystals.
    }
    \label{fig:2measurements}
\end{figure}

We begin the experiment by placing the first HWP in Fig.~\ref{fig:setup}(b) at $0^{\circ}$ and
the second at $67.5^{\circ}$ so that the two crystals implement a
measurement of $\pmb{\pi}_{H}$ followed by $\pmb{\pi}_{D}$, two incompatible
observables. By switching the first HWP to $22.5^{\circ}$ and leaving
the second as it is, the crystals now implement the reverse sequence,
$\pmb{\pi}_{D}$ followed by $\pmb{\pi}_{H}$. In both cases, we record the joint
result of the sequence $\langle \pmb{xy}\rangle$.
In Fig.~\ref{fig:2measurements} we plot this joint result as a function of the input system state
angle, $\theta$. The two orderings
agree within errors. Thus, as expected since the measurement disturbance
is now minimized, the joint result does not depend on the time-ordering
of the measurements. 

So far, nothing surprising has been revealed: when the measurement
disturbance is minimized, the result of a sequence of quantum measurements
is indeed time-ordering invariant. However, there are fundamental phenomena
in quantum physics that only appear in sequences of three or more
measurements. An example is the violation of the Leggett-Garg inequality~\cite{leggett1985quantum,goggin2011violation}. Hence,
we extend the sequence to three measurements, where the third measurement
is a conventional (i.e., `strong') measurement of $\pmb{\Pi}_{K} = \ket{K}\bra{K}$ as implemented by a HWP and PBS (here the capital pi indicates a strong measurement). Our goal is again to test the role of measurement-order when the
disturbance is not a factor. Since this added conventional measurement
will substantially disturb the system and, thus, any subsequent measurements,
we always perform it last. 

\begin{figure}[h!]
    \centering
    \includegraphics[scale=0.45]{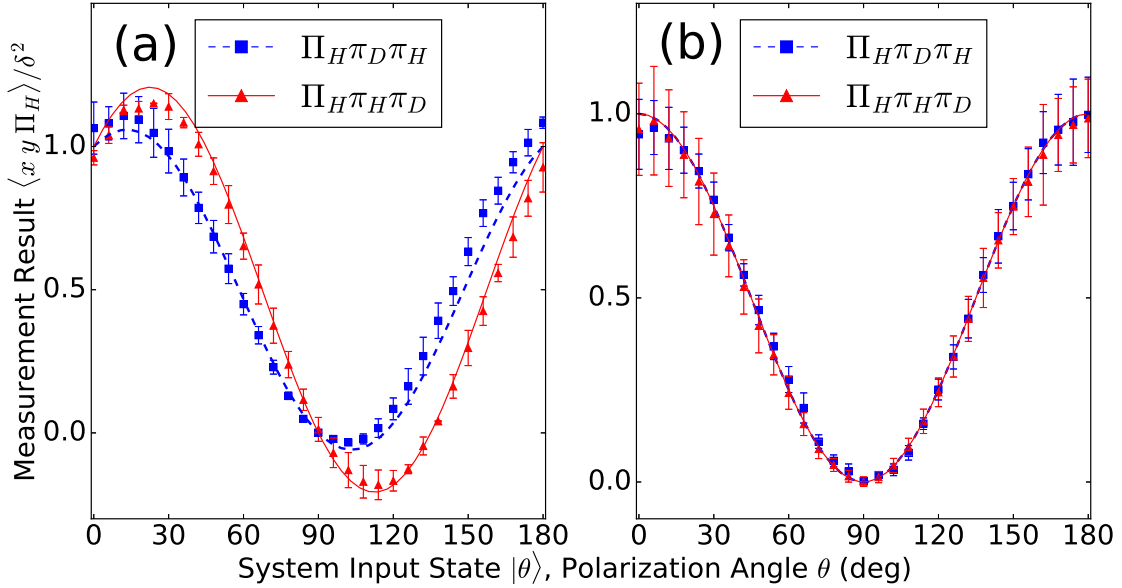}
    \centering
    \caption{The measurement result $\langle \pmb{xy\Pi}_H \rangle/\delta^2$ for a sequence of three measurements. In one, the sequence $\pmb{\Pi}_H \pmb{\pi}_D \pmb{\pi}_H$ is performed (blue squares), and in the other, $\pmb{\Pi}_H \pmb{\pi}_H \pmb{\pi}_D$ is performed (red triangles). The blue dashed and red solid lines are the respective theoretical curves. $\pmb{(a)}$ The input system state is $\ket{\theta}$. The two distinct curves show that the order in which the measurements are made changes the measurement result. $\pmb{(b)}$ An incoherently polarized system state $\rho(\theta)$ is used instead. Now the measurement result does not depend on the ordering of the measurements.}
    \label{fig:Resultstogether}
\end{figure}

The final joint-result of the three measurement sequence is

\begin{equation} \label{eqn:threemeasurement}
\begin{split}
 \langle \pmb{\pi}_K \pmb{\pi}_J \pmb{\pi}_{I}\rangle = \frac{1}{\delta^2} \langle \pmb{\Pi}_{K} \pmb{x}^{(\pmb{\pi}_J)}\pmb{y}^{(\pmb{\pi}_I)}\rangle  = \frac{1}{\delta^2} \int xy\mathrm{Prob}(x,y,K)\mathrm{d}x\mathrm{d}y
\end{split}
\end{equation}
where $\mathrm{Prob}(x,y,K)$ is the probability that a given input photon is transmitted through
the PBS and is detected at transverse coordinate $(x,y)$ on the imaging camera. Since the last is a strong measurement, in Eq.~\ref{eqn:threemeasurement} we directly evaluate the measurement outcome $\langle \pmb{\Pi}_K \rangle$, rather than use the Von Neumann formalism. We set $K = H$ for all of the following measurements. 
Figure~\ref{fig:Resultstogether}(a) shows experimental
results for two orderings of a sequence of three measurements,
$\pmb{\Pi}_{H}\pmb{\pi}_{D}\pmb{\pi}_{H}$ and $\pmb{\Pi}_{H}\pmb{\pi}_{H}\pmb{\pi}_{D}$. The joint
result of one ordering substantially disagrees with the other ordering. Theoretically, the difference is maximum at $\theta = $45$^{\circ}$ and $\theta = $135$^{\circ}$. The nearest experimental points, 48$^{\circ}$ and 138$^{\circ}$, differ by 0.4 $\pm\; 0.1$ and 0.7 $\pm\; 0.1$, respectively.  
Strikingly, quantum physics is not invariant to the time-ordering
even though there is no obvious physical mechanism, such as measurement-induced
disturbance, for this invariance.

Perhaps, while minimized, the residual disturbance still causes this time-ordering asymmetry. But, if this is the case, how could it possibly manifest in three measurements but not two? Specifically, the disturbance would necessarily need to propagate through the first two measurements to reach the third. Hence, one would expect two measurements would be time-asymmetric as well. A recent theoretical investigation offers some mathematical insight ~\cite{diosi2016structural}. In it, our expectation value of $N$ pointer positions $\langle \pmb{x}_1^{(\pmb{A}_1)} \pmb{x}_2^{(\pmb{A}_2)} \cdots \pmb{x}_N ^{(\pmb{A}_N)} \rangle$ was found to be proportional to the recursively nested anti-commutator structure $\{ \{ \dots \{ \{\pmb{A}_N,\pmb{A}_{N-1} \}, \pmb{A}_{N-2} \}, \dots \}, \pmb{A}_1 \}$. While all the anti-commutators are symmetric under interchange of their two arguments, the only anti-commutator that is invariant under interchange of the measurements is the innermost one, which is non-nested. As such, it, and thus the expectation value, is invariant to the ordering of the last two measurements, $\pmb{A}_N$ and $\pmb{A}_{N-1}$. This explains why a sequence of two measurements always exhibits ordering invariance; the entire sequence is the last two measurements. The result of a sequence of three or more measurements will be invariant solely to the ordering of the last two measurements.

While we now have a mathematical reason for the time-ordering dependence of minimally invasive measurements, a physical explanation is still absent. A distinguishing feature of weak measurements is that they preserve the coherence of the measured system and the coherence of the pointer. This can allow a disturbance to propagate in unexpected ways, as in measurement back-action~\cite{PhysRevLett.74.2405,PhysRevA.85.012107}.  Since in a classical system this coherence would be absent, we test what happens to the time-ordering dependence as we decrease the coherence in the H-V basis of the initial system state. To generate these reduced coherence (i.e., mixed) states, we send our polarized input state
$\ket{\theta}$ through a rapidly spinning HWP that is sandwiched
between two quarter-waveplates (QWP).
Since this spinning is faster than the imaging camera acquisition time, the resulting
state is effectively mixed: $\rho(\theta)=\sin^{2}{(\theta)}\ket{H}\bra{H}+\cos^{2}{(\theta)}\ket{V}\bra{V}$. We test the same pair of three-measurement sequences with these
input states. The experiment results, shown in Fig.~\ref{fig:Resultstogether}(b), show that the joint result
of the sequence does not depend on the order in which the observables
were measured. Quantum coherence indeed appears to play an important role in time-ordering symmetry.

To summarize our experimental findings, in the case of two measurements,
the order in which weak measurements are made does not impact the
end result. But, in the case of three or more measurements, the order of the
measurements matter. In short, we have shown that reducing the disturbance induced by measurements does not restore the time symmetry of quantum evolution, as exhibited by the Schrodinger equation. Our findings confirm a recent mathematical description of sequential weak measurements~\cite{diosi2016structural}. While the physical mechanism for this time-ordering invariance is still not clear, we have shown that coherence plays a role. We expect these results will guide the development of closely related areas, such as whether different times of system can be considered separate Hilbert spaces~\cite{Horsman20170395}, and how cause and effect can be identified in quantum systems~\cite{ried2015quantum}.

We would like to thank E. Giese for his thoughtful discussions. This work was supported by the Natural Sciences and Engineering
Research Council (NSERC), the Canada Excellence
Research Chairs (CERC) Program, and the Canada First Research Excellence Fund (CFREF).

\bibliography{refs}

\end{document}